# Extraction of the Schottky parameters in metal-semiconductor-metal diodes from a single current-voltage measurement

Ryo Nouchi

Nanoscience and Nanotechnology Research Center, Osaka Prefecture University, Sakai 599-8570, Japan

E-mail: r-nouchi@21c.osakafu-u.ac.jp

ABSTRACT: In order to develop a method to extract the parameters of the two inherent Schottky contacts from a single current-voltage ($I$-$V$) characteristic curve, the $I$-$V$ characteristics of metal-semiconductor-metal (MSM) diodes with asymmetric Schottky barrier heights are theoretically investigated using the thermionic emission model. The MSM diode structure is commonly used because an additional MS interface is required for the electrical characterization of MS diodes. A finite charge-injection barrier is generally formed at the additional interface. When a local maximum was detected in the first-order derivative of the measured $I$-$V$ characteristics for a MSM diode, the parameters for the Schottky contacts, the zero-bias barrier heights of both MS interfaces, the series resistance of the MSM diode and the effective ideality factor for the MS diode with a higher barrier could be extracted using this method.

## I. INTRODUCTION

The operation of electronic devices such as field-effect transistors, solar cells and electroluminescent diodes is largely governed by the metal-semiconductor (MS) interfaces where charge carriers are injected/extracted. The MS interface can be classified into two types: Schottky and ohmic. If the transport of an electric current through a Schottky interface is hindered by the presence of an energy barrier (Schottky barrier), the current-voltage (*I-V*) characteristics display rectification behavior. Alternatively, ohmic contacts possess (effectively) no energy barrier and their *I-V* characteristics obey Ohm's law. Because of their current rectifying behavior, Schottky-type MS interfaces are employed as a basic electronic component, called a Schottky diode. The current transport through a MS interface with a Schottky barrier is generally treated the same as thermionic emission over the energy barrier. It is expressed as:[1]

$$\begin{aligned} J(V_\mathrm{D}) &= \left(\frac{4\pi q m^* k^2}{h^3}\right) T^2 \exp\left(-\frac{q\Phi_\mathrm{B}}{kT}\right)\left[\exp\left(\frac{qV_\mathrm{D}}{nkT}\right)-1\right] \\ &\equiv A^* T^2 \exp\left(-\frac{q\Phi_\mathrm{B}}{kT}\right)\left[\exp\left(\frac{qV_\mathrm{D}}{nkT}\right)-1\right] \\ &\equiv J_\mathrm{S}\left[\exp\left(\frac{qV_\mathrm{D}}{nkT}\right)-1\right] \end{aligned} \qquad (1)$$

where $q$ is the unit electronic charge, $m^*$ is the effective mass of the charge carrier, $k$ is the Boltzmann constant, $h$ is the Planck constant, $T$ is the absolute temperature, $\Phi_\mathrm{B}$ is the Schottky barrier height in volts, and $V_\mathrm{D}$ is the potential drop across the Schottky diode. The diode expressed by Eq. 1 allows an electric current to flow with positive $V_\mathrm{D}$ values (forward bias), while blocking current with negative $V_\mathrm{D}$ values (reverse bias). Most Schottky diodes deviate from having ideal thermionic emission behavior, which is characterized by a dimensionless parameter called the ideality factor, *n*. This parameter is equal to 1 for thermionic emission and becomes larger than 1 when mechanisms other than thermionic emission, such as field-enhanced tunneling and thermally assisted tunneling, contribute to the current transport. In undoped organic semiconductors, $n$ has been predicted to be larger than 1.2 (ref. 2). $A^*$ is the effective Richardson constant. $J_S$ is the reverse saturation current density and is equal to the absolute current density with a high reverse bias, $|J(-\infty)|$, since the quantity inside the square brackets in Eq. 1 becomes $-1$ when $V_D$ approaches minus infinity. The series resistance ($R_S$) mainly arises from the bulk resistance of the semiconductor, thus $V_D = V - IR_S$, where $V$ is the applied voltage and $I$ is the electrical current. Eq. 1 can be rewritten as:

$$J(V) = J_S \left[ \exp\left(\frac{q(V - IR_S)}{nkT}\right) - 1 \right]. \tag{2}$$

Extraction of the Schottky diode parameters ($\Phi_B$, $n$ and $R_S$) from a single experimentally obtained $I$-$V$ curve was reported using Eq. 2 (refs. 3–7). However, in actual systems, an additional metal contact on the semiconductor is required to exploit the electric current through the Schottky diode. A metal-semiconductor-metal (MSM) structure is necessary to evaluate the MS interface. To observe purely Schottky-type behavior, the additional MS interface on the semiconductor should be ohmic. However, it is difficult to obtain a barrier-free MS interface and a finite energy barrier remains in most Schottky diodes.

In this paper, the $I$-$V$ characteristics of a MSM diode are examined in order to develop a method to extract the parameters of the two inherent Schottky contacts from a single $I$-$V$ characteristic curve. If the difference in the Schottky barrier heights between the two MS interfaces ($\Delta\Phi_B$) is large enough, then the smaller barrier can be ignored and Eq. 2 can be used to examine the experimentally obtained $I$-$V$ characteristics. However, if $\Delta\Phi_B$ is small, the deviation from Eq. 2 becomes significant and proper extraction steps should be followed. In particular, it is important to detect the peak in the $dJ/dV$-$V$ curve, to extract the height of the lower Schottky barrier in addition to the higher barrier.

## II. RESULTS AND DISCUSSION

Figure 1 is a schematic of the MSM diode, where two Schottky diodes are connected back-to-back in series. Conduction of only a single type of charge carriers, either electrons (Fig. 1(a)) or holes (Fig. 1(b)), is considered in the present study. $V_{D1}$ and $V_{D2}$ indicate the voltage drops across the right diode under a reverse bias (Diode 1) and the left diode under a forward bias (Diode 2), respectively. $\Phi_{B1}$ ($\Phi_{B2}$) is the Schottky barrier height of Diode 1 (Diode 2) in volts. From Eq. 1, $J$ can be written as:

$$J = A^* T^2 \exp\left(-\frac{q\Phi_{B1}}{kT}\right)\left[1 - \exp\left(-\frac{qV_{D1}}{nkT}\right)\right] \equiv J_{S1}\left[1 - \exp\left(-\frac{qV_{D1}}{nkT}\right)\right], \quad (3)$$

$$J = A^* T^2 \exp\left(-\frac{q\Phi_{B2}}{kT}\right)\left[\exp\left(\frac{qV_{D2}}{nkT}\right) - 1\right] \equiv J_{S2}\left[\exp\left(\frac{qV_{D2}}{nkT}\right) - 1\right], \quad (4)$$

where $J_{S1}$ and $J_{S2}$ are the reverse saturation current densities for Diodes 1 and 2, respectively. From Eqs. 3 and 4, the voltage drops across each diode, $V_{D1}$ and $V_{D2}$, respectively become:

$$V_{D1} = -\frac{nkT}{q}\ln\left(1 - \frac{J}{J_{S1}}\right), \quad (5)$$

$$V_{D2} = \frac{nkT}{q}\ln\left(1 + \frac{J}{J_{S2}}\right). \quad (6)$$

Here, $V_{MSM}$ is defined as the summation of $V_{D1}$ and $V_{D2}$ ($V_{MSM} \equiv V - IR_S$) and thus $J$ can be rewritten using Eqs. 5 and 6 as (see Appendix for the derivation process):

$$J = \frac{2J_{S1}J_{S2}\sinh\left(\frac{qV_{MSM}}{2nkT}\right)}{J_{S1}\exp\left(-\frac{qV_{MSM}}{2nkT}\right) + J_{S2}\exp\left(\frac{qV_{MSM}}{2nkT}\right)}. \quad (7)$$

Similar I-V characteristics for the MSM diodes have been derived by several groups.[8-12] Herein, the characteristics are further examined by considering the first- and second-order derivatives of $J$ with respect to $V_{MSM}$, which can be written from Eq. 7 as (see Appendix for the derivation process):

$$\{J(V_{MSM})\}' = \frac{q}{nkT}\frac{J_{S1}J_{S2}(J_{S1} + J_{S2})}{\left[J_{S1}\exp\left(-\frac{qV_{MSM}}{2nkT}\right) + J_{S2}\exp\left(\frac{qV_{MSM}}{2nkT}\right)\right]^2}, \quad (8)$$

$$\{J(V_{MSM})\}'' = \left(\frac{q}{nkT}\right)^2 J_{S1}J_{S2}(J_{S1}+J_{S2}) \frac{\left[J_{S1}\exp\left(-\frac{qV_{MSM}}{2nkT}\right) - J_{S2}\exp\left(\frac{qV_{MSM}}{2nkT}\right)\right]}{\left[J_{S1}\exp\left(-\frac{qV_{MSM}}{2nkT}\right) + J_{S2}\exp\left(\frac{qV_{MSM}}{2nkT}\right)\right]^3}$$

$$= A^*\left(\frac{q}{nk}\right)^2 J_{S1}J_{S2}(J_{S1}+J_{S2}) \frac{\left[\exp\left(-\frac{q(2n\Phi_{B1}+V_{MSM})}{2nkT}\right) - \exp\left(-\frac{q(2n\Phi_{B2}-V_{MSM})}{2nkT}\right)\right]}{\left[J_{S1}\exp\left(-\frac{qV_{MSM}}{2nkT}\right) + J_{S2}\exp\left(\frac{qV_{MSM}}{2nkT}\right)\right]^3}.$$

(9)

The local maximum of the first-order derivative appeared at $V_{MSM}$ with $n(\Phi_{B2} - \Phi_{B1}) \equiv n\Delta\Phi_B$, where the second-order derivative became zero.

Figure 2(a) compares the *I-V* curves for MSM diodes with $\Phi_{B2} = 0.5$ V and $\Phi_{B1} = 0.2$ V with MS diodes with various $\Phi_B$ values from 0.2 to 0.8 V. In both cases, $n$ was set to 2.0; they were tested at room temperature ($T = 300$ K) and the $A^*$ value for free electrons (120 A cm$^{-2}$ K$^{-2}$) was used. For $\Phi_B = \Phi_{B2}$, the *I-V* curves become partly identical. From the corresponding *J'* and *J"* curves in Fig. 2(b) and 2(c), the identical region ended at $V_{MSM}$, which is the first local maximum for *J"* and was slightly lower than the local maximum for *J'*. At this point, $V_{D1}$ started to increase (see Fig. 2(d)). In the identical region, the MSM diode could be treated as an MS diode with a single Schottky barrier of $\Phi_{B2}$. Therefore, the Schottky parameters of the apparent MS diode ($\Phi_{B2}$, $n$ and $R_S$) could be extracted by following the reported procedures.[3-7] Finally, the remaining parameter for the MSM diode ($\Phi_{B1}$) could be determined from the local maximum in the first-order derivative of the current (density) using the relationship, $V_{MSM} = n(\Phi_{B2} - \Phi_{B1})$. Fig. 2(e) shows the *J-V* curves for a forward-biased MS diode with $\Phi_B = 0.5$ V and a reverse-biased diode with $\Phi_B = 0.2$ V. The maximum corresponded to the point where the two *J-V*-curves intersected. At the intersection point, the *J* values calculated with Eqs. 3 and 4 were identical.

$$\exp\left(-\frac{q\Phi_{B1}}{kT}\right)\left[1 - \exp\left(-\frac{qV_{D1}}{nkT}\right)\right] = \exp\left(-\frac{q\Phi_{B2}}{kT}\right)\left[\exp\left(\frac{qV_{D2}}{nkT}\right) - 1\right]. \quad (10)$$

The conditions $V_{D1} \gg nkT/q$ and $V_{D2} \gg nkT/q$ were fulfilled at the intersection point in Fig. 2(e). The functions in the square brackets in Eq. 10 can be approximated with:

$$1 - \exp\left(-\frac{qV_{D1}}{nkT}\right) \approx 1,$$
$$\exp\left(\frac{qV_{D2}}{nkT}\right) - 1 \approx \exp\left(\frac{qV_{D2}}{nkT}\right). \tag{11}$$

By using these approximations and taking the natural logarithm of both sides of Eq. 10, the relationship $V_{D2} \approx n(\Phi_{B2} - \Phi_{B1}) = n\Delta\Phi_B$ was obtained. At the intersection point, the condition $V_{D1} \ll V_{D2} \approx V_{MSM}$ held in the MSM diodes, leading to:

$$V_{MSM} \approx n(\Phi_{B2} - \Phi_{B1}). \tag{12}$$

In the derivation above, the inevitable effect of the mirror image of a point charge (a charge carrier in front of the electrode) was not included. The attractive interaction between a point charge and its image induced on the electrode lowers the charge-injection barrier. This lowering, known as the image-force or the Schottky effect,[13] causes the original barrier height ($\Phi_B$) to be replaced by an effective barrier height ($\Phi_B^{eff}$):[14]

$$\Phi_B^{eff} \approx \Phi_B - \delta\Phi_{if}^0 + \left(1 - \frac{1}{n_{if}}\right)V_D \equiv \Phi_B^0 + \left(1 - \frac{1}{n_{if}}\right)V_D, \tag{13}$$

where $\Phi_B^0$ is the zero-bias barrier height. $\delta\Phi_{if}^0$ is the zero-bias image-force lowering:

$$\delta\Phi_{if}^0 = \left[\frac{2q^2 N}{(4\pi)^2 \varepsilon_\infty^2 \varepsilon_s^3 \varepsilon_0^3} q\left(\left|V_{bi}^0\right| - kT\right)\right]^{1/4}, \tag{14}$$

where $N$ is the ionized impurity concentration; $\varepsilon_\infty$ and $\varepsilon_s$ are the optical and static dielectric constants for the semiconductor, respectively; $\varepsilon_0$ is the permittivity of a vacuum; and $V_{bi}^0$ is the zero-bias built-in potential (the interface band-bending) in volts. $n_{if}$ is the ideality factor, written as:

$$n_{if} = \left(1 - \frac{\delta\Phi_{if}^0}{4\left|V_{bi}^0\right|}\right)^{-1}, \tag{15}$$

which describes the bias dependence of the barrier heights of ideal Schottky diodes (pure thermionic emission) caused by the effect of the image-force. Also, $n_{if}$ is different from $n$, which depends on contributions from the charge transport processes other than thermionic emission.[1] Eq. 1 can be rewritten by replacing $\Phi_B$ with $\Phi_B^{\text{eff}}$ as:

$$J(V_D) = A^*T^2 \exp\left(-\frac{q\Phi_B^0}{kT}\right)\exp\left[-\frac{qV_D}{kT}\left(1-\frac{1}{n_{\text{if}}}\right)\right]\left[\exp\left(\frac{qV_D}{nkT}\right)-1\right]. \quad (16)$$

For $V_D \gg nkT/q$, Eq. 16 can be approximated as:

$$J(V_D) \approx A^*T^2 \exp\left(-\frac{q\Phi_B^0}{kT}\right)\exp\left(\frac{qV_D}{n^{\text{eff}}kT}\right),$$

$$n^{\text{eff}} \equiv \frac{nn_{\text{if}}}{n+n_{\text{if}}-nn_{\text{if}}}. \quad (17)$$

In the same voltage region, the original expression, Eq. 1 can be approximated as:

$$J(V_D) \approx A^*T^2 \exp\left(-\frac{q\Phi_B}{kT}\right)\exp\left(\frac{qV_D}{nkT}\right)$$

$$\ln[J(V_D)] \approx \frac{q}{nkT}V_D + \ln(A^*T^2) - \frac{q\Phi_B}{kT}. \quad (18)$$

$\Phi_B$ and $n$ can be extracted from the $y$-intercept and the slope of the $\ln|J|$-$V_D$ curve, respectively. By comparing Eqs. 17 and 18, it can be understood that including the image-force effect changes the extracted values for $\Phi_B$ and $n$ to $\Phi_B^0$ and $n^{\text{eff}}$, respectively. The maximum position in the $J'$-$V$ characteristic curve occurred at the point where the $J$-$V$-curve for the forward-biased MS Diode 2 and that of the reverse-biased MS Diode 1 intersect. By applying this procedure to Eqs. 16 and 17, we obtained the relationship:

$$V_{\text{MSM}} \approx n_2^{\text{eff}}\left(\Phi_{B2}^0 - \Phi_{B1}^0\right) \equiv n_2^{\text{eff}}\Delta\Phi_B^0, \quad (19)$$

at the intersection point. The suffix for $n^{\text{eff}}$ indicates the diode number. The effective ideality factor for Diode 2 ($n_2^{\text{eff}}$) was acquired by linearly fitting the forward-biased region with the identical region shown in Fig. 2(a). Therefore, the $n_2^{\text{eff}}$ in Eq. 19 is the same as the $n^{\text{eff}}$ for an MS diode with a higher barrier ($n_{\text{high}}^{\text{eff}}$). Thus, Eq. 19 becomes:

$$V_{\text{MSM}} \approx n_{\text{high}}^{\text{eff}}\left(\Phi_{B2}^0 - \Phi_{B1}^0\right) \equiv n_{\text{high}}^{\text{eff}}\Delta\Phi_B^0. \quad (20)$$

To observe how the image-force effect alters the shape of the $J$-$V$ curves, silicon MS diodes were examined. The physical constants for silicon are as follows: $m^* = 0.33\ m_0$ (for electrons; $m_0$ is the free electron mass) and $\varepsilon_\infty = \varepsilon_s = 11.9$. The $J$-$V$ characteristics with and without the image-force effect are compared in Fig. 3(a) for $T = 300$ K, $\Phi_B = 0.5$ V, $n = 1.2$, $N = 10^{16}$ cm$^{-3}$ and $V_{bi}^0 = 0.1$ V. The saturation behavior observed in the reverse-biased region was significantly weakened by bias-induced lowering of the barrier height. Using the dataset, the zero-bias barrier height and the effective ideality factor could be extracted by linearly fitting the $\ln|J|$-$V_D$ curves in the voltage region $V_D \gg nkT/q$ (Eqs. 17 and 18). These parameters extracted by the fitting procedure were the same as the $\Phi_B^0$ ($\equiv \Phi_B - \delta\Phi_{if}^0$) calculated using Eq. 14 and the $n^{eff}$ calculated with Eq. 17. The calculated values for these two quantities are shown in Fig. 3(b) and 3(c) for low ($10^{14}$ cm$^{-3}$), moderate ($10^{16}$ cm$^{-3}$) and high ($10^{18}$ cm$^{-3}$) dopant concentrations. Using the calculated $n^{eff}$ value shown in Fig. 3c, the $V_{MSM}$ at the (local) maximum in the $J'$-$V_{MSM}$ curves could be obtained using Eq. 20. Figure 3d shows the calculated results for $N = 10^{16}$ cm$^{-3}$ with lower barrier heights ($\Phi_{B1}^0$) of 0.2, 0.3, 0.4 and 0.5 V. The results reflect the trend for $n^{eff}$. Thus, MSM diodes with high $N$ and/or low $V_{bi}^0$ values (generally, corresponding to a low $\Phi_B$) require high voltages to obtain a peak in the $J'$-$V$ characteristics.

## III. CONCLUSION

In summary, the $I$-$V$ characteristics of MSM diodes with asymmetric Schottky barrier heights were theoretically investigated and a method to extract the parameters of the two inherent Schottky contacts was proposed. These diodes are important for characterizing common MS diodes, since a barrier-free MS interface is needed to electrically characterize ideal MS diodes, which are difficult to obtain. First, analytical expressions for the $J$-, $J'$- and $J''$-$V$ characteristics were derived based on the thermionic emission model. The voltage corresponding to the peak in the $J'$-$V$ curve was found to be equal to the product of the ideality factor and the difference between the two barrier heights. Next, the image-force effect, which is inevitable in MS diodes, was included and the same relationship between the voltage corresponding to the local maximum in the $J'$-$V$ curve and the product of the two quantities was found to hold. It was necessary to replace the two quantities with the effective ideality factors and the difference between the two zero-bias barrier heights.

The Schottky parameters for the two MS interfaces could be extracted from a single $I$-$V$ characteristic curve by following the procedures below: (1) the $R_S$ for the MSM diode can be extracted by following the reported fitting procedures, such as the Werner's method.[6,7] The fitting range of $|V|$ must be lower than the voltage at the first local maximum in the $I''$-$V$ curve, which is slightly lower than the local maximum in the corresponding $I'$-$V$ curve. (2) $V$ should be corrected to become $V_{MSM} \equiv V - IR_S$ using the extracted $R_S$. (3) The $\varPhi_B^0$ and $n^{eff}$ values for a MS diode with the higher barrier can be extracted by linearly fitting the corrected $\ln|I|$-$V_{MSM}$ curve. The fitting range of $|V_{MSM}|$ must be lower than the voltage at the first local maximum in the $I''$-$V_{MSM}$ curve, which is slightly lower than the local maximum in the $I'$-$V_{MSM}$ curve. In addition, in the fitting range for $V_{MSM}$, the condition $|V_{MSM}| \gg nkT/q$ should be fulfilled to ensure that there is a linear region in the $\ln|I|$-$V_{MSM}$ curve. (4) The $\varPhi_B^0$ for the MS diode with a lower barrier can be calculated with Eq. 20, using the $V_{MSM}$ at the peak position in the $I'$-$V_{MSM}$ curve. If a local maximum is detected in the $I'$-$V$ curves for the MSM diodes, the parameters for the Schottky contacts, the $\varPhi_B^0$ values for the two MS interfaces, the $R_S$ of the MSM diode and the $n^{eff}$ for the MS diode with a higher barrier can be acquired from a single $I$-$V$ characteristic curve.

**ACKNOWLEDGMENTS**

This work was supported in part by the Special Coordination Funds for Promoting Science and Technology from the Ministry of Education, Culture, Sports, Science and Technology of Japan and by a Grant-in-Aid for Challenging Exploratory Research (No. 25600078) from the Japan Society for the Promotion of Science.

**APPENDIX: DERIVATION OF EXPRESSIONS FOR MSM DIODES**

As described in the main text, the voltage drops across each MS diode in a MSM diode become:

$$V_{D1} = -\frac{nkT}{q}\ln\left(1-\frac{J}{J_{S1}}\right),$$

$$V_{D2} = \frac{nkT}{q}\ln\left(1+\frac{J}{J_{S2}}\right).$$

The voltage drop across the MSM diode, $V_{MSM}$ is defined as the summation of $V_{D1}$ and $V_{D2}$:

$$V_{MSM} = V_{D1} + V_{D2} = \frac{nkT}{q}\left[\ln\left(1+\frac{J}{J_{S2}}\right) - \ln\left(1-\frac{J}{J_{S1}}\right)\right],$$

$$\frac{qV_{MSM}}{nkT} = \ln\left(\frac{1+\dfrac{J}{J_{S2}}}{1-\dfrac{J}{J_{S1}}}\right),$$

$$\exp\left(\frac{qV_{MSM}}{nkT}\right) = \frac{1+\dfrac{J}{J_{S2}}}{1-\dfrac{J}{J_{S1}}} = \frac{J_{S1}J_{S2} + J_{S1}J}{J_{S1}J_{S2} - J_{S2}J}.$$

By solving this expression for $J$, we obtain Eq. 7 as:

$$J = \frac{J_{S1}J_{S2}\exp\left(\dfrac{qV_{MSM}}{nkT}\right) - J_{S1}J_{S2}}{J_{S1} + J_{S2}\exp\left(\dfrac{qV_{MSM}}{nkT}\right)}$$

$$= \frac{J_{S1}J_{S2}\exp\left(\dfrac{qV_{MSM}}{2nkT}\right) - J_{S1}J_{S2}\exp\left(-\dfrac{qV_{MSM}}{2nkT}\right)}{J_{S1}\exp\left(-\dfrac{qV_{MSM}}{2nkT}\right) + J_{S2}\exp\left(\dfrac{qV_{MSM}}{2nkT}\right)}$$

$$= \frac{2J_{S1}J_{S2}\sinh\left(\dfrac{qV_{MSM}}{2nkT}\right)}{J_{S1}\exp\left(-\dfrac{qV_{MSM}}{2nkT}\right) + J_{S2}\exp\left(\dfrac{qV_{MSM}}{2nkT}\right)}.$$

To obtain the first-order derivative of $J$ with respect to $V_{MSM}$, Eq. 8, we introduce auxiliary functions as:

$$J \equiv \frac{f(V_{MSM})}{g(V_{MSM})},$$

$$f(V_{MSM}) = J_{S1}J_{S2}\exp\left(\frac{qV_{MSM}}{2nkT}\right) - J_{S1}J_{S2}\exp\left(-\frac{qV_{MSM}}{2nkT}\right),$$

$$g(V_{MSM}) = J_{S1}\exp\left(-\frac{qV_{MSM}}{2nkT}\right) + J_{S2}\exp\left(\frac{qV_{MSM}}{2nkT}\right),$$

$$\{f(V_{MSM})\}' = \frac{qJ_{S1}J_{S2}}{2nkT}\left[\exp\left(\frac{qV_{MSM}}{2nkT}\right) + \exp\left(-\frac{qV_{MSM}}{2nkT}\right)\right],$$

$$\{g(V_{MSM})\}' = \frac{q}{2nkT}\left[J_{S2}\exp\left(\frac{qV_{MSM}}{2nkT}\right) - J_{S1}\exp\left(-\frac{qV_{MSM}}{2nkT}\right)\right].$$

By using these functions, Eq. 8 can be obtained as follows:

$$\{J(V_{MSM})\}' = \frac{\{f(V_{MSM})\}'g(V_{MSM}) - f(V_{MSM})\{g(V_{MSM})\}'}{[g(V_{MSM})]^2}$$

$$= \frac{qJ_{S1}J_{S2}}{2nkT}\frac{1}{[g(V_{MSM})]^2}$$

$$\times \left\{\left[\exp\left(\frac{qV_{MSM}}{2nkT}\right) + \exp\left(-\frac{qV_{MSM}}{2nkT}\right)\right]\left[J_{S1}\exp\left(-\frac{qV_{MSM}}{2nkT}\right) + J_{S2}\exp\left(\frac{qV_{MSM}}{2nkT}\right)\right]\right.$$

$$\left. - \left[J_{S2}\exp\left(\frac{qV_{MSM}}{2nkT}\right) - J_{S1}\exp\left(-\frac{qV_{MSM}}{2nkT}\right)\right]\left[\exp\left(\frac{qV_{MSM}}{2nkT}\right) - \exp\left(-\frac{qV_{MSM}}{2nkT}\right)\right]\right\}$$

$$= \frac{qJ_{S1}J_{S2}}{2nkT}\frac{1}{[g(V_{MSM})]^2}$$

$$\times \left\{J_{S1} + J_{S2}\exp\left(\frac{qV_{MSM}}{nkT}\right) + J_{S1}\exp\left(-\frac{qV_{MSM}}{nkT}\right) + J_{S2}\right.$$

$$\left. - \left[J_{S2}\exp\left(\frac{qV_{MSM}}{nkT}\right) - J_{S2} - J_{S1} + J_{S1}\exp\left(-\frac{qV_{MSM}}{nkT}\right)\right]\right\}$$

$$= \frac{q}{nkT}\frac{J_{S1}J_{S2}(J_{S1} + J_{S2})}{\left[J_{S1}\exp\left(-\frac{qV_{MSM}}{2nkT}\right) + J_{S2}\exp\left(\frac{qV_{MSM}}{2nkT}\right)\right]^2}.$$

The second-order derivative of $J$ with respect to $V_{MSM}$, Eq. 9 can be derived as:

$$\{J(V_{\text{MSM}})\}'' = \left(\frac{q}{nkT}\right)^2 J_{S1} J_{S2} (J_{S1} + J_{S2}) \frac{\left[J_{S1} \exp\left(-\frac{qV_{\text{MSM}}}{2nkT}\right) - J_{S2} \exp\left(\frac{qV_{\text{MSM}}}{2nkT}\right)\right]}{\left[J_{S1} \exp\left(-\frac{qV_{\text{MSM}}}{2nkT}\right) + J_{S2} \exp\left(\frac{qV_{\text{MSM}}}{2nkT}\right)\right]^3}.$$

By substituting the expressions of Eqs. 3 and 4 respectively into $J_{S1}$ and $J_{S2}$ in the square brackets of the numerator, the second-order derivative can be rewritten as:

$$\{J(V_{\text{MSM}})\}'' = A^* \left(\frac{q}{nk}\right)^2 J_{S1} J_{S2} (J_{S1} + J_{S2}) \frac{\left[\exp\left(-\frac{q(2n\Phi_{B1} + V_{\text{MSM}})}{2nkT}\right) - \exp\left(-\frac{q(2n\Phi_{B2} - V_{\text{MSM}})}{2nkT}\right)\right]}{\left[J_{S1} \exp\left(-\frac{qV_{\text{MSM}}}{2nkT}\right) + J_{S2} \exp\left(\frac{qV_{\text{MSM}}}{2nkT}\right)\right]^3}.$$

Figures

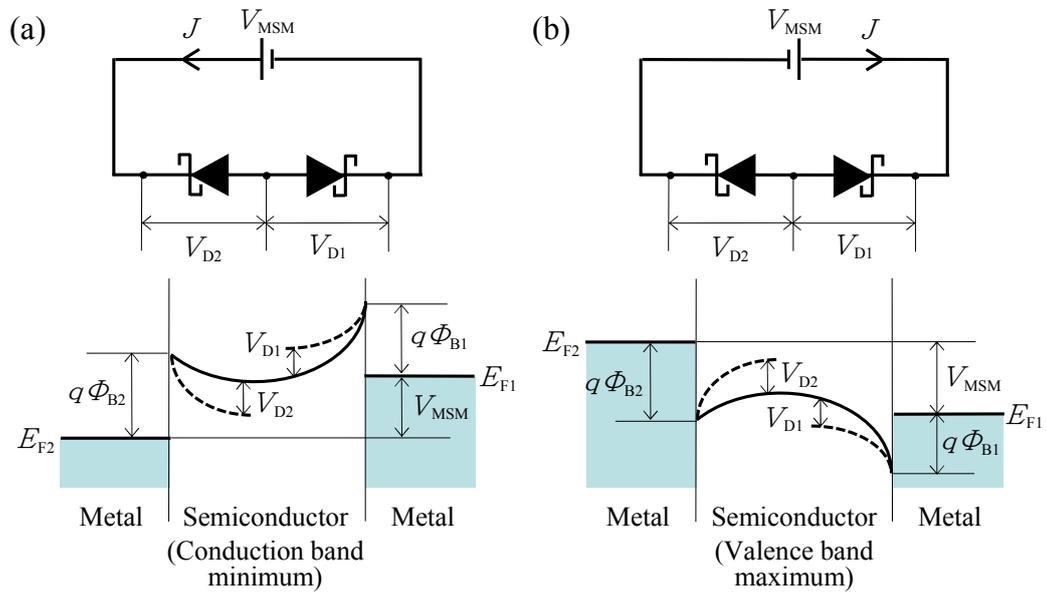

FIG. 1. Schematic diagrams for MSM diodes with (a) electron and (b) hole conduction.

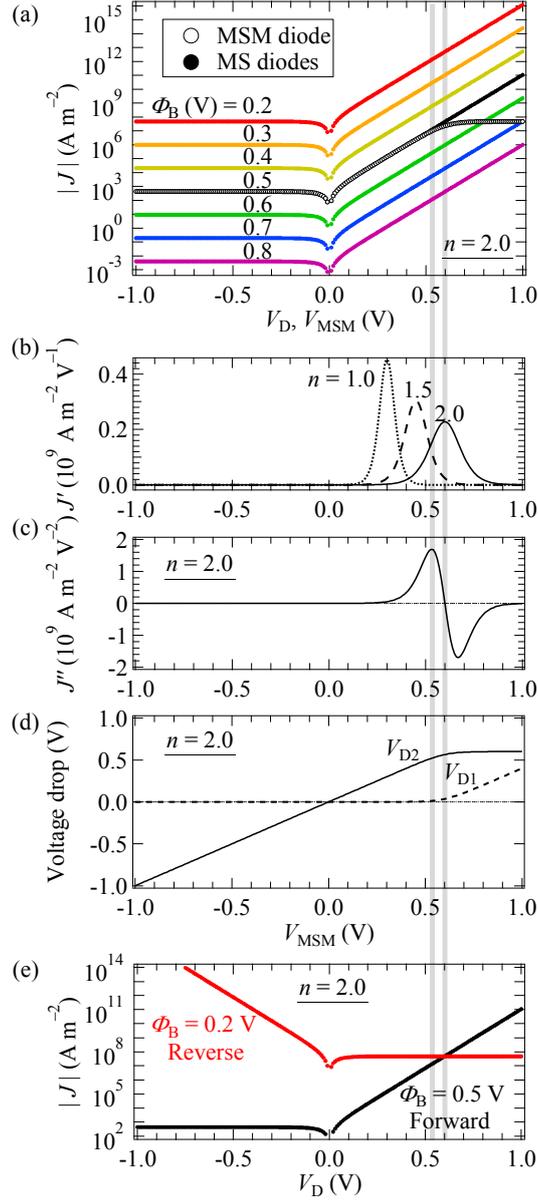

FIG. 2. Characteristics of the MS and MSM diodes without the image-force effect for $T = 300$ K, $A^* = 120$ A cm$^{-2}$ K$^{-2}$ and $n = 2.0$. (a) $J$-$V_D$ curves for MS diodes with various $\Phi_B$ values from 0.2 to 0.8 V (calculated by Eq. 1), along with the $J$-$V_{MSM}$ curves for a MSM diode with $\Phi_{B2} = 0.5$ V and $\Phi_{B1} = 0.2$ V (calculated by Eq. 7). Corresponding $V_{MSM}$ dependences of (b) $J'$, (c) $J''$ and (d) the voltage drops across each MS diode in the MSM diode in (a). (e) $J$-$V_D$ curves for a forward-biased MS diode with $\Phi_B = 0.5$ V and a reverse-biased one with $\Phi_B = 0.2$ V (calculated by Eq. 1). The vertical gray lines indicate the point where the $J$-$V_{MSM}$ curve for the MSM diode started to deviate from the $J$-$V_D$ curve for the MS diode in (a) and the maximum point of the $J'$-$V_{MSM}$ curve in (b).

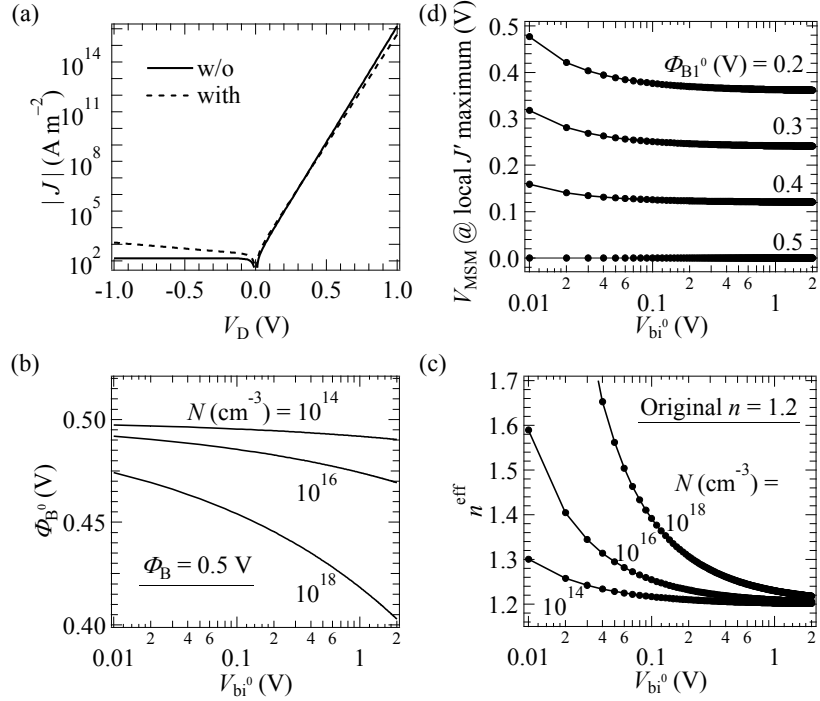

FIG. 3. The image-force effect on the characteristics of the MS and MSM diodes based on electron conduction in silicon for $T = 300$ K and $n = 1.2$. (a) $J$-$V_D$ curves for MS diodes with and without the image-force effect for $\Phi_B = 0.5$ V, $N = 10^{16}$ cm$^{-3}$ and $V_{bi}^0 = 0.1$ V. (b) The zero-bias barrier height and (c) the effective ideality factor for $\Phi_B = 0.5$ V. (d) $V_{MSM}$ at the local maxima in the $J'$-$V_{MSM}$ curves for $N = 10^{16}$ cm$^{-3}$. The higher barrier height ($\Phi_{B2}$ in this calculation) was set to 0.5 V and various values for the lower barrier height ($\Phi_{B1}$ in this calculation) were examined.

16